\documentclass[manuscript]{aastex}

\shorttitle{The Fundamental Plane of Open Clusters}
\shortauthors{Pang et al.}

\begin{document}


\title{The fundamental plane of open clusters}


\author{Xiaoying Pang\altaffilmark{1,3}}

\affil{Shanghai Institute of Technology, 100 Haiquan Road,
Fengxian district, Shanghai 201418, P.R. China\\
}
 \email{xypang@bao.ac.cn}

\author{Shiyin Shen \altaffilmark{2,3}}

\author{Zhengyi Shao \altaffilmark{2,3}}


\affil{Shanghai Astronomical Observatory, 80 Nandan Road, Shanghai,
  P.R. China\\
\and 
 Key Lab for Astrophysics, Shanghai Normal University, 100 Guilin
Road, Shanghai 200234, P.R. China
}



\begin{abstract}

We utilize the data from the Apache Point Observatory Galactic Evolution Experiment-2 (APOGEE-2) in the fourteenth data release of the Sloan Digital Sky Survey (SDSS) to calculate the line-of-sight velocity dispersion $\sigma_{1D}$ of a sample of old open clusters (age larger than 100\,Myr) selected from the Milky Way open cluster catalog of Kharchenko et al. (2013). Together with their $K_s$ band luminosity $L_{K_s}$, and the half-light radius $r_{h}$ of the most probable members, we find that these three parameters show significant pairwise correlations among each other. Moreover, a fundamental plane-{\it like} relation among these parameters is found for the oldest open clusters (age older than 1\,Gyr), $L_{K_s}\propto\sigma_{1D}^{0.82\pm0.29}\cdot
r_h^{2.19\pm0.52}$ with $rms \sim\, 0.31$\,mag in the $K_s$ band absolute magnitude. The existence of this relation, which deviates significantly from the virial theorem prediction, implies that the dynamical structures of the old open clusters are quite similar, when survived from complex dynamical evolution to age older than 1 Gyr.

\end{abstract}



\section{Introduction}
In early-type galaxies, there is a
tight relation between the effective radius, the central velocity dispersion, and
the average surface brightness within the effective radius, which is called 
the ``fundamental plane'' (FP; Djorgovski \& Davis 1987, Dressler et al. 1987).  
Djorgovski (1995) extended the FP to another old population, the
Galactic globular clusters (GCs) and found that the FP of GCs at
the core radius agreed well with the virial plane $R_c\propto \sigma^{2.2\pm0.15
}I_0^{-1.1\pm0.1}$, which lives up to the
expectation that the system is old enough to settle into equilibrium. 

Unlike the old, isolated, and massive GCs in the Galactic halo, open clusters (OCs) are 
young clusters in the Galactic plane with mass in the range of $\rm 100-10^4\,M_\odot$ (Binney \&
Merrifield 1998). They are located in the crowded plane of the Milky
Way, where molecular clouds are abundant. 
The encounters between OCs and the interstellar clouds increase 
the internal cluster energy, and consequently 
 lead to the expansion and disruption of the OCs (Spitzer 1958, Kruijssen 2012).  
Many studies have suggested a typical survival timescale of
200\,Myr for OCs (Friel 1995;
Sarajedini et al. 1999; Yang et al. 2013), and  
only 3 percent of the known OCs have ages above 1\,Gyr
(Chumak et al. 2010). 
The surviving OCs avoid disruption by usually having larger mass,  more concentrated density profiles and are in orbits that may avoid the destructive influence of molecular clouds in the disk (Friel 1995).  On the other hand, 
old OCs deviate from simple virial equilibria due to their complex and ``aggressive'' tidal environments, despite their ages being many times of their dynamical timescale,  which would otherwise drive them to a quasi-equilibrium state. 

From observation, Bonatto \& Bica (2005) first tried to derive an FP
for 11 OCs based on parameters of mass (overall cluster mass, core radius and overall mass density). There seemed to be a trend of an FP in OCs, which can be explained by the correlation among cluster mass, radius and density. However, they could not quantitatively draw any conclusions due to low number statistics and lack of kinematic data.  
There are several obstacles referring to an accurate estimation of the FP of
OCs. 

First of all, the membership of stars is poorly determined with
photometry alone. In many cases, the available kinematic data of OCs, neither proper motions nor
radial velocities (RVs) are precise or homogeneous enough to guarantee a
secure discrimination between field and cluster stars.
Additionally, velocity dispersions cannot be properly calculated with such
poor kinematic data. 
Without a reliable list of members, the derivation of total brightness 
is also affected. The incompleteness of faint stars makes it even worse. 
Last but not least, the morphology of OCs, as the name 
``open'', suggests, is kind of
irregular and is not as spherical as GCs. Finding out the center of
OCs is painstaking, making even the size estimate of OCs from model fitting quite poor (Seleznev 1994).

Nowadays, the quality of the data for OCs is greatly improved. 
Kharchenko et al. (2013, 2016) compiled a catalog containing more than 3000 OCs with uniform
photometry from 2MASS (Skrutskie et al. 2006). 
The size and luminosity of OCs
  estimated by Kharchenko et al. (2013, 2016) are measured at near infrared wavelengths ($J,\, H, \,K_s$), where  the Galactic extinction is minimized (Mathis 1990). 
Additionally, the high-resolution
spectral survey of Galactic stars (e.g., APOGEE-2), allows 
the measurement of internal velocity dispersion of OCs. Therefore, the time is ripe to revisit the FP of OCs, extending the FP to low-mass systems.

This paper is organized as follows. The Kharchenko catalog and 
SDSS/DR14 APOGEE-2 data are introduced in Section 2. 
Measurements of the structural parameters and the velocity dispersion are
described in Section 3. The exploration of the FP of OCs is 
presented in Section 4. Finally, we  make a brief summary and discussion in Section 5.

\section{Open cluster data and sample}

The OC catalog of Kharchenko et al. (2013) is based on the $J$,
$H$, $K_s$ band photometry from 2MASS data (Skrutskie et al. 2006), plus proper motions taken from
Roeser et al. (2010). 
Here we list the essential aspects referring to this paper. The central coordinates of OCs were taken as the points of maximum surface density of the most
probable cluster members.  The member selection procedure was only applied to stars located within the apparent radius $r_2$, which is the distance from the cluster center to where the 
cluster stellar density is equal to that of the field. 
Photometric probabilities $P_{JH}$ and $P_{JK}$ were computed via the
color magnitude diagram (CMD), while kinematic
probabilities were calculated based on proper motions. 
The most probable ($1\sigma$) members were defined as members with both photometric
and kinematic probabilities larger than $61\%$. The ages of OCs were 
determined via isochrone fitting to the most probable
 members in the central part by using the turn-off stars as cluster age indicators  (Kharchenko et al. 2013). Kharchenko et al. (2016) computed the 
intrinsic integrated $J$, $H$, $K_s$ magnitudes for OCs in the catalog
of Kharchenko et al. (2013) by integrating the observed luminosity profiles with
corrections for incomplete faint members.

The APOGEE-2 from the fourteenth data release (DR14) of SDSS (Abolfathi et al. 2018) is a high resolution ($\sim22500$) and high
S/N ($>$100) spectroscopic survey in the infrared wavelength range of
1.51-1.70\,$\rm \mu m$, which has observed hundreds of Galactic OCs (Majewski et
al. 2016). 
The major observed stellar objects of APOGEE-2 were red
giant branch stars and red clump stars, with some bright 
main-sequence stars as well.  
The accuracy of the RV of giants reached a level of
0.125\,$\rm km\,s^{-1}$ (S/N$>20$, Nidever et al. 2015), which is not only sufficient to
discriminate the cluster members from field stars, but also to the study of the internal dynamics of OCs.

Young OCs are still partially embedded in their parental
 molecular clouds, adding complication to the analysis of their
 dynamics. 
Therefore, in this study, we only take OCs older than 100\,Myr in Kharchenko et al. (2013) catalog, which we call Sample I. 
We then select stars from APOGEE-2, which are located within
the apparent radii from the centers of OCs in Sample I. We only keep stars with both [Fe/H] and RV measurements and then only keep the OCs with more than 10 stars inside the apparent radius. This selection results in 153 OCs totally. Not all selected stars are members of OCs. To maximize the utilization of the kinematic information, we do not count on the memberships of the Kharchenko catalog here. Rather, for each OC, we check the [Fe/H]--RV diagram of the selected stars by eye (see Appendix), and exclude OCs that do not show apparent peaks in the diagrams (i.e. not enough member stars). This is done to ensure a reliable measurement of the internal velocity dispersion of OCs (see Section 3.2 for more details). 

Finally, we get 26 OCs for velocity dispersion measurements (age $\ge100$\,Myr), 18 of them with age older than 1\,Gyr. These 26 old OCs are generally brighter compared to the original 153 OCs in the distribution of $K_s$ band luminosity and half-light radius (panel a in Figure 1). This is a bias due to the fact that the observed targets of APOGEE-2 are mainly red giants. However, there is no difference between the 26 and 18 OCs (panel b in Figure 1, Kolmogorov-Smirnov test with p=0.98).

\begin{figure}[h]
\includegraphics[width=9cm]{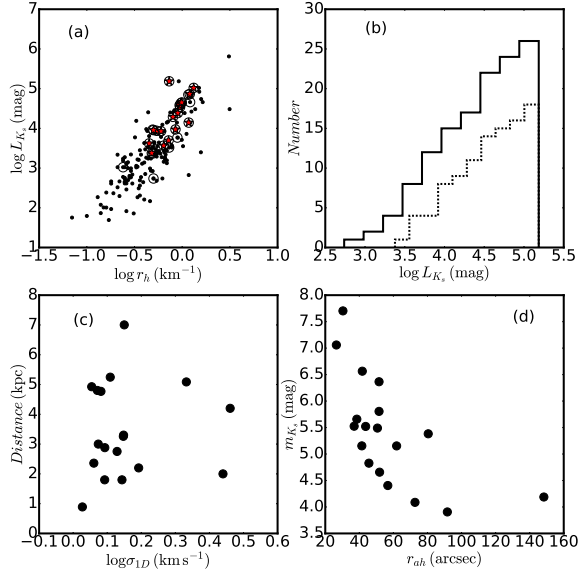}
\centering
\caption{Panel a: The distribution of $K_s$ band luminosity $L_{K_s}$ and half-light radius $r_h$. Black dots are the 153 OCs after the first cross-match between APOGEE-2 and Kharchenko catalog. Open circles are the 26 OC sample. Red stars are the 18 old OCs used in the FP study.
Panel b: the cumulative distribution of $L_{K_s}$ for the 26 good OCs (solid line) and 18 old OCs (dotted line). 
Distributions of old OCs are presented in panel c (distance verses velocity dispersion $\sigma_{1D}$), and d (apparent magnitude $m_{K_s}$ verses the apparent radius $r_{ah}$).
 \label{Fig.1} }
\end{figure}

\section{Properties of Open Clusters}
\subsection{Photometric and Structural parameters}

We take the $K_s$ band absolute magnitude $M_{K_s}$ from Kharchenko et al (2016), which has already been corrected for Galactic reddening and unseen faint stars. However, no error of $M_{K_s}$ is available for individual OCs in the Kharchenko catalog. 
The error budget of $M_{K_s}$ is comprised of three uncertainties: the integrated apparent magnitude $m_{K_s}$, the distance and the reddening. According to Kharchenko et al. (2012), the typical uncertainty of distance modulus is 0.35\,mag, and  the uncertainty of extinction is  $\sim0.06\,$mag. The error in $m_{K_s}$ is mainly due to random errors in the magnitudes of member stars (also called stochastic effect) and the incomplete counting of faint stars, whereas the former term is the dominant one since the total flux of a star cluster is dominated by giant stars.  We use the bootstrap technique to estimate this term of error. Specifically, for each OC, we construct 100 simulated OCs by bootstrapping its most probable members. We compute the integrated apparent magnitudes $m_{K_s}'$ of 100 realizations and take the standard deviation as the error of $m_{K_s}$. Finally, we combine the errors in $m_{K_s}$, distance modulus and reddening for each OC.

Considering the diverse profile of OCs, we take a model-independent measurement, the half-light radius $r_{h}$, as the structural parameter. Specifically, we use the most probable members in Kharchenko catalog and count the summed flux of member stars inside-out until the radius where the flux is half of the total. We also use the bootstrap realizations to estimate the error in $r_h$. Here, not all the member stars but only the visible ones are used for the calculation of $r_h$. The mass segregation effect (Pang et al. 2013) might bias the $r_h$ we measure. To test this effect, we also count the half-number radius, the radius where the number of most probable member stars is half the total, and find that it is almost indistinguishable from the $r_h$ of each OC. 
The error of $r_h$ is quite small at the order of 0.01-0.05\,pc.

\subsection{The line-of-sight velocity dispersion}

The accuracy of RVs allows us to compute the line-of-sight velocity
dispersion ($\sigma_{1D}$) among member stars. We use stars having high S/N and RV errors less than 0.125 km/s. Typically, these stars of high S/N also have [Fe/H] measurements. To further reduce possible biases from membership determination, we fit the [Fe/H] and RV distribution with a two-dimensional Gaussian mixture model $\Phi_{tot}$, which is constituted of two components, the cluster $\Phi_c$ and the field $\Phi_f$. Under this assumption,  
the likelihood of a star with given parameters is $\mathcal{L}_i=\Phi_i$,

\begin{equation}
\Phi_{i}=n_c\cdot\Phi_{c,i}+(1-n_c)\cdot\Phi_{f,i}. 
\end{equation}
\begin{equation}
\Phi_{c,i}=\mathcal{N}(RV_i | RV_{c}, \sigma'_{1D,ci})\cdot \mathcal{N}(M_i| M_{c}, \sigma'_{m,ci}),
\end{equation}
\begin{equation}
\Phi_{f,i}=\mathcal{N}(RV_i | RV_{f}, \sigma'_{1D,fi})\cdot \mathcal{N}(M_i| M_{f}, \sigma'_{m,fi}),
\end{equation}

where $n_c$ is the fraction of the cluster component, $\mathcal{N}(*)$ is Gaussian distribution function.
$\sigma'_{1D,ci}$, $\sigma'_{m,ci}$, $\sigma'_{1D,fi}$ and $\sigma'_{m,fi}$ are the dispersions of $\mathcal{N}(*)$. $RV_c$ ($RV_f$) is the mean radial velocity of cluster (field), and $M_c$ ($M_f$) is the mean metallicity of cluster (field), respectively. 
Note that $\sigma'_{1D,i}$ and $\sigma'_{m,i}$ are not constants, but composed of two terms, the intrinsic dispersion $\sigma_{1D,*}$ ($\sigma_{m,*}$) and the individual observational uncertainty $er_{RV,i}$ ($er_{m,i}$), which varies one by one star. 
\begin{equation}
\sigma_{1D(m),i}^{'2}=\sigma_{1D(m),*}^{2}+er_{RV(m),i}^2
\end{equation}

Parameters are fitted so that the total likelihood of all stars $\mathcal{L}={\displaystyle \prod_{i=1}^{N}} \mathcal{L}_i$ reaches maximum. We use nested sampling to derive the probability density function (PDF) of parameters. Since the resulted PDF is almost Gaussian, we take the mean and standard deviation of the PDF, as the best estimation of the parameters and their errors. 

As an example in Figure 2, the blue dots show the [Fe/H] and RV values for individual selected stars in NGC\,1245, where the two-dimensional Gaussian mixture model (Equation 1) is projected into one-dimension to show how well the model fits the data. The [Fe/H]--RV distribution is fitted well by the sum of field (green lines) and member (red lines) components.

 The members of OCs observed in APOGEE-2 are spatially random sampling of the OC members. Therefore, we do not expect this random member list will bias the computation of the velocity dispersion of OCs. Typically, the RV measurements of more distant OCs have lower S/N. If our $\sigma_{1D}$ measurement were biased by the RV uncertainties, we would expect that $\sigma_{1D}$ of more distant OCs are biased to higher values. To further test whether the result would be biased by the precision of RV measurements, we show the derived $\sigma_{1D,c*}$ as a function of distance (Pearson correlation test: r=0.07, p=0.77) in panel c of Figure 1. 
 The independence of $\sigma_{1D,c*}$ on distance, further confirms that our measurement of velocity dispersion is not affected by the observational uncertainty. 

\begin{figure}[h]
\includegraphics[width=8cm]{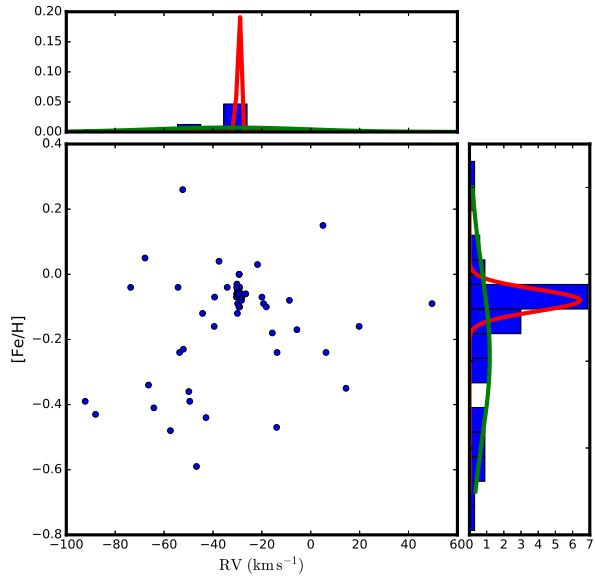}
\centering
\caption{The distribution of [Fe/H] and RV (blue dots) for the open cluster NGC\,1245 is fitted
  with a two-dimensional Gaussian model. For illustration, we project
  this model into one dimension. The
  distributions of [Fe/H] (right histogram) and RV (top 
  histogram) are constituted of two components,
  the field (green lines) and the cluster (red lines). The histogram is normalized such that the integral over the range is 1. 
\label{Fig.1} }
\end{figure}

\section{The fundamental plane of open clusters}

Figure 3 shows the pairwise correlations among the above three global parameters of 26 OC sample (old OCs: black dots; young OCs: open circles), where $r$ and $s$ are the Pearson and Spearman's rank correlation coefficients, respectively. The probability of the null hypothesis ($p$) of each correlation test is also shown.

\begin{figure}[h]
\includegraphics[width=11cm]{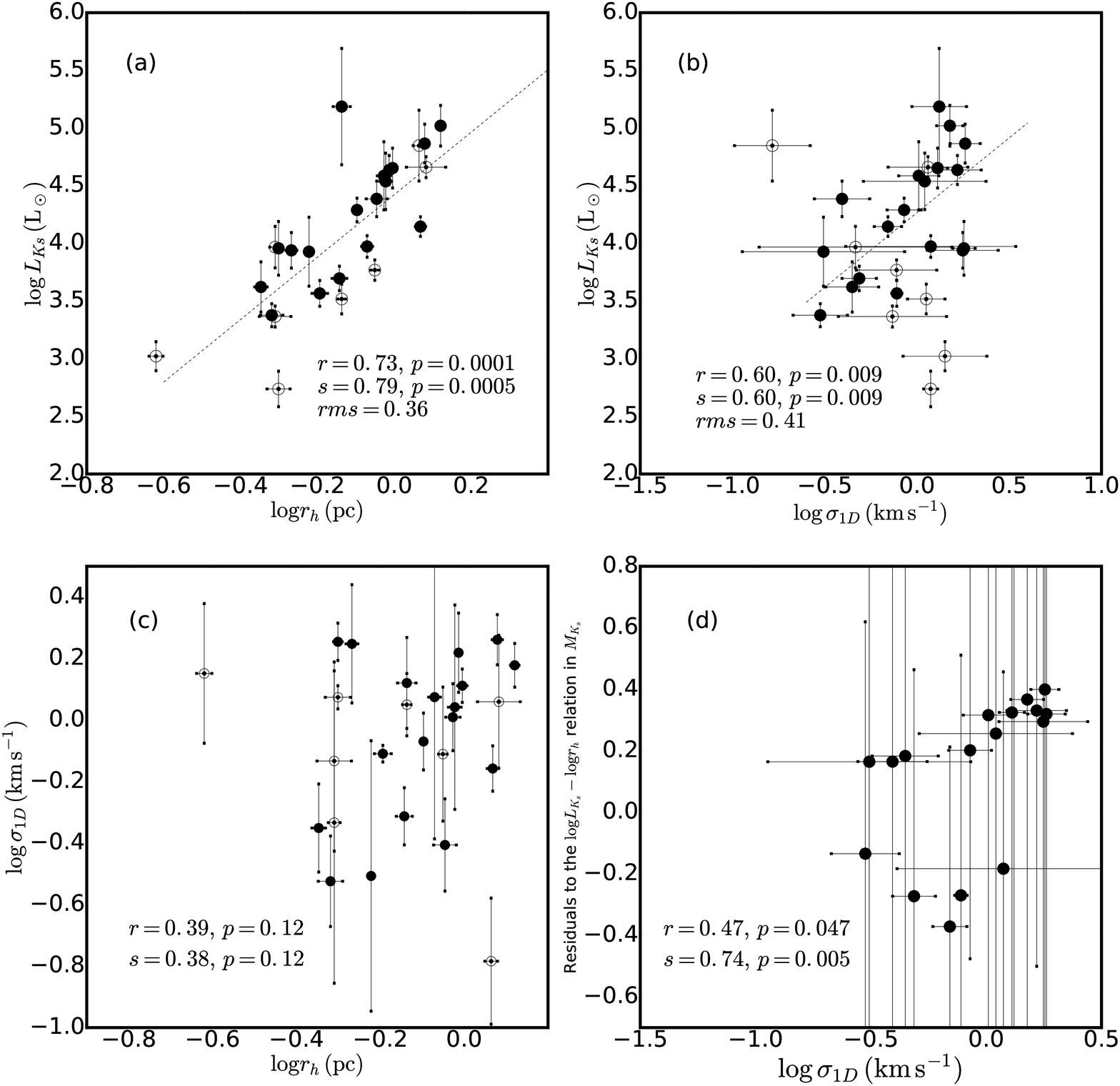}
\centering
\caption{The pairwise correlations (panels a-c) among the luminosity in the $K_s$ band $L_{K_s}$, the line-of-sight velocity dispersion $\sigma_{1D}$, and the half-light radius $r_h$ for old OCs (age$\,>\,$1\,Gyr, black dots) and OCs younger (100\,Myr\,$<$\,age$\,<\,$1\,Gyr, open circles). The Pearson ($r$) and Spearman's  rank ($s$) correlation coefficients, and the p-value ($p$) are shown in each panel (for old OCs only). The dotted line is the fitted linear relation for old OCs weighted with errors of both X and Y axes (panel a and b). Panel d shows the dependence of individual residuals in $M_{K_s}$ from the best fitting linear relation of $L_{K_s}$--$r_h$ (dotted line in panel a) on the velocity dispersion for old OCs.  
\label{Fig.1} }
\end{figure}

Generally, the luminosity $L_{K_s}$ increases monotonically with both the half-light radius $r_{h}$ (panel a) and the velocity dispersion $\sigma_{1D}$ (panel b). These two correlations become even tighter for the subsample of old OCs (age$>$1\,Gyr, black dots). 
The strongest correlation is $L_{K_s}$--$r_h$ relation, which has $r=0.73$ and $s=0.79$ (for the old OCs). There are two possible origins of this tight correlation. One is that the old OCs have similar density profiles so that the larger OCs are also brighter. The other possibility is the distance bias, i.e. the OCs at larger distance are always brighter and larger. To test whether there exists a distance bias in our OC sample, we also show the relation between the apparent magnitude and apparent size (in unit of arcsec) in panel d of Figure 1. As can be seem, these two apparent quantities still show a strong correlation, which confirms that the correlation shown in panel a of Figure 3 is induced from the similar density profiles of OCs.

We perform linear regression for two strongest relations: the $L_{K_s}$--$r_h$ and $L_{K_s}$--$\sigma_{1D}$ for the 18 old OCs.  The coefficients of the linear relations are computed with least-square regression weighted by errors, which are combinations of errors on X and Y axes. We also have checked possible covariances between the errors of each set of parameters. The Pearson correlation tests all suggest very weak covariances (all have $p>0.5$).

The best fitting results are 
\begin{equation}
{\rm log}L_{K_s}=(2.71\pm0.56)\cdot {\rm log}r_{h}+(4.42\pm0.10),
\end{equation}
and 
\begin{equation}
{\rm log}L_{K_s}=(1.30\pm0.39)\cdot {\rm log}\sigma_{1D}+(4.27\pm0.10).
\end{equation}
These two relations are plotted as the dotted lines in panel a and b of Figure 3 and the fitting coefficients are also listed in Table 1. In terms of $M_{K_s}$, the $rms$ (root-mean scatter) of these two fitting relations are 0.36 and 0.41 mag respectively.

In analogy with GCs, if old OCs were also in dynamical equilibrium, we would expect an even tighter relation among $L_{K,s}$, $\sigma_{1D}$, and $r_{h}$, i.e. an FP of OCs. Before exploring this possibility, we calculate the residual of the $L_{K_s}$--$r_h$ relation, which indeed shows a significant and positive correlation with $\sigma_{1D}$ ($r$=0.47, $s$=0.74), having p-values smaller than 0.05 ($p_r$=0.047, $p_s$=0.005). That is to say, for OCs at a given radius, the brighter OCs would also have higher $\sigma$. Generally, this behavior is consistent with the idea that higher mass galaxies would have higher kinetic energy when they were in dynamical equilibrium.

To make a more quantitative parametrization of the FP of OCs, we fit the relation
\begin{equation}
{\rm log}L_{K_s}=a\cdot {\rm log}\sigma_{1D}+b\cdot {\rm log} r_{h}+c,
\end{equation}
in the three dimensional parameter space and obtain $a=0.80\pm0.29$, $b=2.18\pm0.52$, and $c=4.45\pm0.08$ using multivariate weighted least-square model. We combine uncertainties of $L_{K_S}$, $\sigma_{1D}$ and $r_h$ as  weights in the regression.

\begin{figure}[h]
\includegraphics[width=15cm]{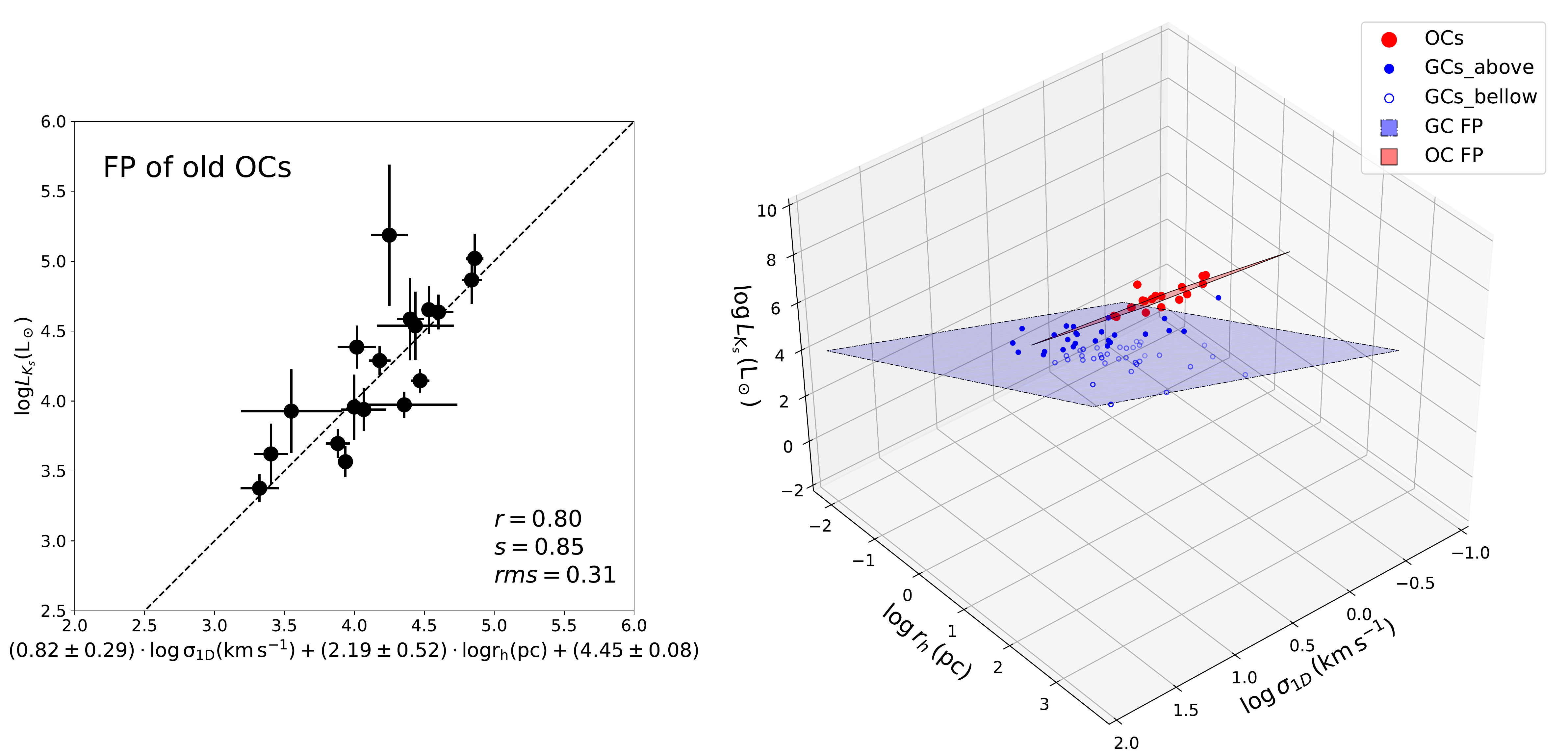}
\centering
\caption{Left: the multivariate correlation for the luminosity in the $K_s$ band $L_{K_s}$, the line-of-sight velocity dispersion $\sigma_{1D}$, and the half-light radius $r_{h}$. This shows an edge-on view of the FP of old OCs. The coefficients of the fitted correlation (dashed line) are shown below the figure. The regression is performed on the black circles, weighted by their errors in both X and Y. Right: The FP of old OCs (red plane) and GCs at core radius (blue plane). Red points are the 18 old OCs in our sample, blue dots (above the blue plane) and circles (below the blue plane) are GCs taken from Harris et al. (2010). An animation of this 3D plot with different viewing angles are available for online version.
\label{Fig.2} }
\end{figure}

We present an edge-on view of the FP of the old OCs in the left panel of Figure 4. For the resulted FP of OCs, the $rms$ in $M_{K,s}$ is 0.31\,mag, which is significantly smaller than that of the $L_{K_s}$--$r_h$ (0.36 mag) and $L_{K_s}$--$\sigma_{1D}$ relations (0.41\,mag, Table 1). The reduced $\chi^2$ (4.09\,mag, number of degrees of freedom is 15, Table 1) is also reduced from the bivariate relations. 
Thus confirms a plane-{\it like} relation exists in the three-dimensional space of (${\rm log}L_{K_s}, {\rm log}\sigma_{1D}, {\rm log}r_h$). The scatter of the FP may be partly explained by stochastic effects, i.e. a stochastic and under-sampling of the initial mass function (Piskunov et al.s 2011, Anders et al. 2013) due to the low mass of OCs (few $\rm 10^3\,M_\odot$). Parameters of these 18 old OCs are presented in Table 2. We cross check the age, distance, reddening of the 18 old OCs of Kharchenko et al. (2013) with other references (Dias et al. 2002, Bhattacharya et al. 2017). Generally, all these parameters agree with reference values within the typical errors (Kharchenko et al. 2012). 
 None of these 18 OCs overlap the 11 OCs in Bonatto \& Bica (2005), where the first FP of OCs were estimated based on overall mass, core radius and projected overall mass density.

The FP of old OCs we obtained can be approximated by $L_{K_s} \propto \sigma_{1D}r_h^2$, which shows significant deviation from the virial theorem $L_{K_s} \propto \sigma_{1D}^2r_h$. 
 This large deviation implies a complicated dynamical status of OCs.

 First, $L_{K_s}$ is roughly  proportional to $\sigma_{1D}$ rather than follows the $L_{K_s} \propto \sigma_{1D}^2$  relation, which means that $\sigma_{1D}$ is larger than the virial theorem prediction at a given $L_{Ks}$ and implies that the old OCs are still expanding.  
When the OCs formed, the gas dispersed, causing OCs to expand (Kroupa et al. 2001, Pfalzner et al. 2014, Brinkmann et al. 2017, Shukirgaliyev et al. 2017, Kuhn et al. 2018) and resulting in a vulnerable shallow potential. 
Usually the loose and low-mass OCs will disrupt within a few 100\,Myrs and merge into the Galactic disk (Lada \& Lada 2003; Gouliermis 2018). 
Only the massive OCs with high star formation efficiency can stubbornly go through Gyrs of dynamical evolution, fighting against the tidal field destruction and cluster expansion, and eventually disrupt at the timescale of Gyr (Shukirgaliyev et al. 2018).

On the other hand, $L_{K_s}$ is approximately proportional to $r_{h}^2$ (Equation 5 \& 7). This relation tends to suggest a constant surface brightness for old OCs, which might be caused by both the dynamical evolution and selection effects. Dynamical heating drives the high surface brightness OCs to expand, while the selection effects may keep the low surface brightness OCs from being selected. N-body simulations indeed show that OCs above 1\,Gyr have larger star formation efficiency and higher density contrast than young OCs, exhibiting similar compact density profile (Shukirgaliyev et al. 2018).

In the right panel of Figure 4, we over-plot the FP (blue plane) of GCs derived at core radius ($L\propto r^{1.07}\sigma^{1.67}$, Djorgovski 1995), which resembles virial plane ($L\propto r^{1}\sigma^{2}$) very much. Blue dots are observed GCs taken from Harris (2010), where solid/open ones show GCs above and below the FP respectively. As can be seeen, all the old OCs (red dots, red plane) are above the FP of GCs, locating in the super-virial region, which is consistent with the expansion scenario. However, we are still unclear why the old OCs form such a tight plane, which might be connected to the combined effects of internal evolution and tidal effect from the Galactic plane (Friel 1995; Lada \& Lada 2003; Gouliermis 2018). Detailed NBODY simulations would be helpful to further clarify its physical implication (Trenti \& van del Marel 2013).

\begin{deluxetable}{ccccc}
\tablecolumns{5}
\tablewidth{0pc}
\tablecaption{Coefficients and rms for the correlations of old OCs}
\tablehead{
\colhead{a} & \colhead{b}  & \colhead{rms in $M_{K_s}$ (mag)} & \colhead{Correlation} & reduce $\chi^2$}
\startdata
   $2.71\pm0.56$ & $\sim$ & 0.36 & ${\rm log}L_{K_s}=a\cdot{\rm log}r_{h}+c$ & 5.19\\      
\hline                        
   $1.30\pm0.39$ & $\sim$ & 0.41 & ${\rm log}L_{K_s}=a\cdot{\rm log}\sigma_{1D}+c$ & 7.12\\      
\hline                        
  $0.80\pm0.29$ & $2.19\pm0.52$ & 0.31 & ${\rm log}L_{K_s}=a\cdot{\rm log}\sigma_{1D}+b\cdot{\rm log}r_{h}+c$ & 4.09\\      
\enddata
\end{deluxetable}

\begin{deluxetable}{ccccccc}
\tablecolumns{7}
\tablewidth{0pc}
\tablecaption{Parameters of the 18 old OCs}
\tablehead{
\colhead{Sequence number} & \colhead{Name}  & \colhead{Age(logt)}  & \colhead{Distance\,(kpc)}  &  \colhead{$\sigma_{1D}\,{\rm (km^{-1})}$} & \colhead{$r_h\,{\rm (pc)}$} & \colhead{${\rm log}L_{K_s}$(mag)}}
\startdata
  42  &          FSR 0494  &  9.30  &  5.09  &  1.10  &  0.95  &   4.54    \\
 146  &            IC 166  &  9.00  &  4.80  &  1.51  &  1.32  &   5.02    \\
 255  &       Berkeley 66  &  9.15  &  7.00  &  0.39  &  0.90  &   4.39    \\
 264  &          NGC 1245  &  9.02  &  3.00  &  0.70  &  1.17  &   4.15    \\
 265  &            King 5  &  9.09  &  2.20  &  1.77  &  0.54  &   3.94    \\
 483  &          NGC 1798  &  9.30  &  5.25  &  1.02  &  0.94  &   4.59    \\
 508  &        Czernik 20  &  9.19  &  2.00  &  0.31  &  0.60  &   3.93    \\
 509  &       Berkeley 17  &  9.60  &  1.80  &  0.45  &  0.45  &   3.62    \\
 634  &       Berkeley 71  &  9.02  &  3.26  &  0.30  &  0.48  &   3.38    \\
 733  &          NGC 2158  &  9.33  &  4.77  &  1.82  &  1.20  &   4.86    \\
 933  &        Trumpler 5  &  9.50  &  2.75  &  1.66  &  0.97  &   4.64    \\
1292  &          NGC 2420  &  9.36  &  2.88  &  0.49  &  0.72  &   3.70    \\
1585  &          NGC 2682  &  9.53  &  0.89  &  0.78  &  0.64  &   3.57    \\
3088  &          NGC 6791  &  9.65  &  4.93  &  1.29  &  0.99  &   4.65    \\
3155  &          NGC 6819  &  9.21  &  2.36  &  1.80  &  0.50  &   3.96    \\
3435  &       Berkeley 53  &  9.09  &  3.30  &  1.32  &  0.73  &   5.19    \\
3655  &       Berkeley 98  &  9.32  &  4.20  &  1.19  &  0.85  &   3.97    \\
3779  &          NGC 7789  &  9.27  &  1.80  &  0.85  &  0.80  &   4.29    \\
\enddata
\begin{flushleft}
Sequence number of OCs, name, age, and distance are taken from Kharchenko et al. (2013) catalog. 
The parameters $\sigma_{1D}$, $r_h$, and ${\rm log}L_{K_s}$ are computed in this paper based on APOGEE-2 and Kharchenko et al. (2013).
\end{flushleft}
\end{deluxetable}

\section{Summary}

In this study, we have combined the kinematic data from the APOGEE-2 of the SDSS/DR14 (Abolfathi et
al. 2017) with the OC sample of Kharchenko et al. (2013, 2016), and obtained a FP-{\it like} relation in the logarithm space of  the luminosity at $K_s$ band
$L_{K_s}$, the line-of-sight velocity dispersion $\sigma_{1D}$, and the half-light radius $r_{h}$ for a sample of 18 dedicatedly selected old OCs (age $>$ 1 Gyr).
The FP of OCs is expressed as ${\rm log}L_{K_s}=(0.80\pm0.29)\cdot {\rm log}\sigma_{1D}+(2.19\pm0.52)\cdot {\rm log} r_{h}+(4.45\pm0.08)$, and the relation is quite narrow with a rms in $M_{ks}\sim$\,0.31\,mag. 
We argue that the FP of OCs is established through their complicated dynamical evolution. Because of their long time-scale evolution, the old OCs show self-similar density profiles. On the other hand, because of the continuous dynamical heating  from the interaction with the Galactic disk, the old OCs are only in quasi-equilibrium state and are still expanding.

Although a self-consistent evolutionary scenario of OCs is implied from the FP we derived, there are still some uncertainties on its specific shape.
First, as we already discussed, the stochastic effect may induce scatter of the FP of OCs. During the FP fitting, our bootstrap error estimation of the luminosity of OCs (section 3.1) has  partly compensated this effect. However, it is unclear about how the the stochastic effect varies with the age and mass of the OCs and whether this variation would cause any systematical bias in the FP fitting. For the velocity dispersion, $\sigma_{1D}$, the high resolution spectroscopy of APOGEE stars and our dedicated selection criteria make the 18 old OCs the few and largest OC sample with accurate $\sigma_{1D}$ measured to date. However, one of the uncertainties that we have not considered is the internal motion from binaries, which will cause $\sigma_{1D}$ to be overestimated (Kouwenhoven \& de Grijs 2008).  

If the binary fraction in OCs does not vary systematically and significantly with their mass, we would expect that the correction for the binarity is roughly a constant. That is to say, the shape of the FP of OCs would not be biased by this effect. Recent studies have suggested that the binary fraction of main sequence stars may depend on the mass of clusters (Milone et al. 2012). To estimate the possible influence of this correlation on our measurements of $\sigma_{1D}$, 
we assume the binaries in our giants follow the same period orbital distribution as that of Raghavan et al. (2010). We assume the maximum period changes from $10^4$ (Geller et al. 2012, Mermilliod et al. 2007) to $10^5$ days for our brightest ($M_{K_s}$=5.2) and faintest ($M_{K_s}$=3.4) OCs. We find a binary fraction varies from 18\% to 48\%. According to Geller et al. (2010), this binary fraction changes will give a correction on $\sigma_{1D}$ about 0.1\,$\rm km\,s^{-1}$, which is still significantly smaller than our typical $\sigma_{1D}$. We therefore do not expect the variation of binary fraction affects our current results.  
 
More importantly, our sample of old OCs is still too small (only 18) and its dynamical range of the  parameters (especially $\sigma_{1D}$) is also too small. The second data release (DR 2) of Gaia mission (Gaia collaboration 2018) is an unprecedented astronomical dataset for OCs. Cantat-Gaudin et al. (2018) established membership for 1212 OCs based on the proper motions and parallaxes of Gaia DR 2. Further
investigations  including Gaia data would  increase the sample size and possibly the dynamical range of the structural/dynamical parameters, which
could  further verify the conclusions in this study.  

Finally, in order to understand the physical process during the dynamical evolution of OCs in more detail, we are comparing the results with numerical simulations to investigate the origin of the FP among old OCs (Pang et al., in preparation). Specific mechanism that is responsible to the FP will be uncovered.

\acknowledgments

This work is funded by National Natural Science Foundation of China,
No: 11503015 \& 11673032  (XP), 11573050 \& 11433003 (SS), and 11390373 (ZS), and 
Project 973 No: 2014CB845705  (SS). XYP is grateful to the travel
grants supported by Sonderforschungsbereich 881 ``The Milky Way System'' of the German
Research Foundation (subproject B5). 

We are gratitude to Prof. Dr. Chenggang Shu for inspiring suggestions. We also give thanks to Prof. Dr. Andreas Just, Bekdaulet Shukirgaliyev, and Dr. Yohai Meiron for helpful
discussions. We are grateful to Shih-Yun Tang for his help in 3D Python plotting, and Fabian Klein for his help in the regular expression realization, and Shuai Feng for his help in MCMC. We thank Dr. Chien-Cheng Lin for his help in retrieving APOGEE data. 

\appendix 

\section{Appendix}
We introduce the procedure of open cluster selection from APOGEE-2 of
SDSS/DR14. The [Fe/H]--RV diagrams are examined by eye. If there
is no concentration of stars present in the [Fe/H]--RV diagram
(Figure 5, upper-left panel), it implies that the number of cluster members is too small and therefore most of these stars are from the field. Even though a few cluster members 
might be observed, their number is not sufficient to generate an over-density in the diagram. We plot the CMD and check the location of the
stars referring to the isochrone at the age of the
cluster (upper-right panel). The OC's age, distance and reddening used in the CMD are taken from Kharchenko et al. (2013). Field stars may be located far from the isochrone, while cluster members right on top of it. Stars located on the isochrone might be members. However, their number is too small to be used for quantitative studies. For comparison, the bottom panels in Figure 5 present an example OC that has enough members and shows an over-density both in the [Fe/H]--RV diagram and on top of the isochrone.

\begin{figure}[h]
\includegraphics[width=10cm]{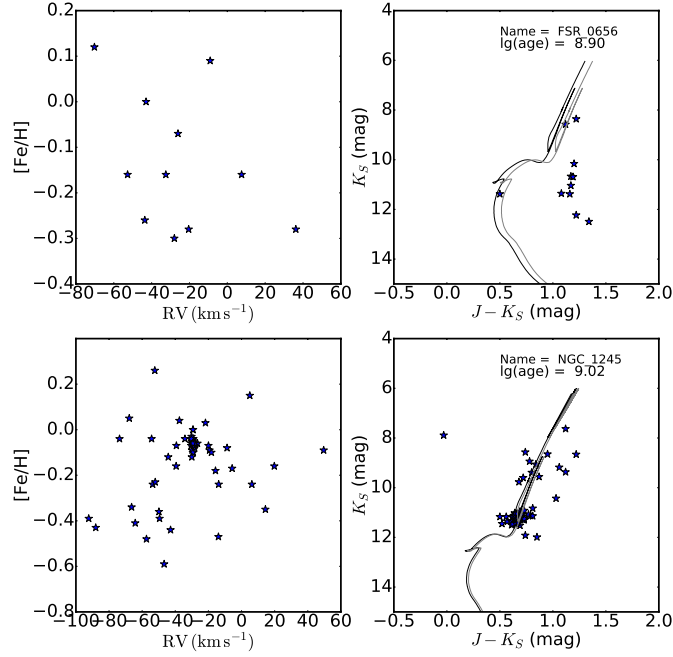}
\centering
\caption{Upper panels: a bad OC sample in our selection. 
The upper-left panel is the [Fe/H]--RV diagram, and the upper-right panel is the CMD. The blue starred symbols
are stars observed from APOGEE-2. The solid curves are isochrones computed from the CMD3.0 on-line server$^0$. 
The black one is 
corrected for the extinction law of Fitzpatrick (1999), while the grey for 
Cardelli et al. (1989). Bottom panels: a good OC sample that meets our selection criteria, which shows an over-density in the [Fe/H]--RV diagram (bottom-left panel) and the CMD (bottom-right panel). 
\label{Fig.1} }
\end{figure}
\footnotetext{http://stev.oapd.inaf.it/cgi-bin/cmd}

\end{document}